\documentclass[12pt]{spieman}  
\usepackage{amsmath,amsfonts,amssymb}
\usepackage{gensymb}
\usepackage{graphicx}
\usepackage{setspace}
\usepackage{tocloft}
\usepackage{lineno}
\usepackage{cleveref}
\usepackage{booktabs}
\usepackage{xcolor}

\newif\ifrevision
\revisionfalse
\ifrevision
  \newcommand{\rev}[1]{{\color{red}#1}}
\else
  \newcommand{\rev}[1]{#1}
\fi

\newcommand\Bbbbone{\boldsymbol{1}}

\title{Understanding HWO's Field of Regard and Characterization Requirement Trade Space with a Dynamic Observation Scheduling Algorithm}

\author[a,b]{Corey Spohn}
\author[a,b]{Christopher C. Stark}
\author[c,d]{Dmitry Savransky}
\author[a,b]{Natasha Latouf}
\affil[a]{NASA Goddard Space Flight Center, Greenbelt, MD 20771, USA}
\affil[b]{Sellers Exoplanet Environments Collaboration, 8800 Greenbelt Road, Greenbelt, MD 20771, USA}
\affil[c]{Cornell University, Ithaca, NY 14853, USA}
\affil[d]{Carl Sagan Institute, Cornell University, Ithaca, NY 14853, USA}

\cftpagenumbersoff{figure}
\cftpagenumbersoff{table} 
\begin{document} 
\maketitle

\begin{abstract}
    The Habitable Worlds Observatory (HWO) aims to image and characterize at
    least 25 ExoEarth candidates (EECs). Achieving this goal requires a
    detailed understanding of the observatory's design trade space, including
    the operational efficiency of the EEC survey. This study quantifies the
    impact of two critical parameters: the instantaneous field of regard (FoR)
    and the number of characterization observations required per EEC
    ($N_\text{char}$). We introduce a novel dynamic scheduling algorithm
    implemented within the EXOSIMS framework that models information gain
    during the mission. The scheduler models the orbital information known
    about each planet and forecasts detection probabilities to make scheduling
    decisions. We explore a multi-dimensional trade space, varying aperture
    size (6.5\,m and 8.0\,m), dedicated EEC survey time (2.5, 5.0, 7.5 years),
    $N_\text{char}$ (1 to 4), and FoR ($15\degree$ to $135\degree$). Our
    results demonstrate that the FoR is a major driver of the mission yield,
    with the yield decreasing significantly when the FoR is less than
    $90\degree$. We find that increasing $N_\text{char}$ imposes a significant
    cost to mission yield, as each additional characterization required reduces
    yield by approximately 22\%. The cumulative impact of requiring four characterizations
    instead of one lowers the yield by approximately 52\%. This harsh penalty can
    be partially mitigated by increasing the survey duration. The relative yield
    loss when increasing $N_\text{char}$ from 1 to 2 is 38\% for a 2.5 year survey
    and 14\% for a 7.5 year survey. Our results highlight the complex interactions
    between HWO's engineering constraints and science requirements, and emphasize that
    the EEC survey efficiency is a critical component of HWO's design space.
\end{abstract}

\keywords{Exoplanets, Direct Imaging, Habitable Worlds Observatory (HWO), Yield Modeling, Scheduling Optimization, Field of Regard}

{\noindent \footnotesize\textbf{*}Corey Spohn, \linkable{corey.a.spohn@nasa.gov} }

\begin{spacing}{2}   

\section{Introduction}
\label{sect:intro}

Early trade studies for direct imaging missions established the foundational
principle of yield modeling, 
evaluating observatory designs by calculating the expected number of planets
the mission will detect by calculating the mission's total ``completeness''
\cite{brownObscurationalCompleteness2004,brownSingleVisitPhotometric2005}.
Completeness represents the fraction of an assumed planet population, such as
exoEarth candidates (EECs), that can be detected by a given observation of
a star. By summing the completeness of all target stars, one can calculate
the expected number of planets that will be detected by the mission,
referred to as the mission's ``yield''. This approach was expanded upon by
Ref.~\citenum{savransky_analyzing_2010} to simulate direct imaging
missions observation-by-observation and led to the direct imaging mission
yield calculator EXOSIMS\cite{delacroix_science_2016,Savransky2015exosims}. This was
followed by Refs.~\citenum{starkMaximizingExoEarth2014,stark_lower_2015} which
established AYO, a probabilistic yield calculator that calculates the
expected yield of a mission by optimizing the integration time allocated to
each target star.

EXOSIMS and AYO were used to create quantitative yield comparisons of observatory designs
during the LUVOIR\cite{TheLUVOIRTeam2019} and
HabEx\cite{gaudiHabitableExoplanet2020} mission concept studies.
The LUVOIR and HabEx studies showed that a direct imaging mission was viable
and informed the Astro 2020 Decadal Survey's recommendation of a space-based
direct imaging mission capable of detecting and characterizing 25 EECs\cite{nationalacademiesofsciences_pathways_2021}.
That recommendation led to NASA's selection of the Habitable Worlds
Observatory (HWO) as a future flagship mission capable of reaching the 25 EEC goal.
Accomplishing this goal will be a significant challenge and requires careful
consideration of the observatory's design trade space to meet the
science requirement while remaining within time and budget constraints.
\rev{Much of the trade space is poorly understood, particularly operational
constraints such as the observatory's instantaneous field of regard (FoR),
the annular region of the sky set by the telescope's allowable Sun-pointing
angles (see \Cref{sub:field_of_regard}). Because the FoR determines when
individual targets are observable, a narrow FoR can create scheduling
bottlenecks that compound over a multi-year survey.}

Many fundamental parameters have been studied extensively. For example,
Ref.~\citenum{stark_lower_2015} showed that the mirror diameter is the
dominant driver of mission yield, with the EEC yield increasing
quadratically with the mirror diameter. However, optimizing the mission
\rev{will also require understanding how operational constraints affect scheduling
efficiency.} HWO will have to balance long
exposure times, the seasonal availability of stars, and the time required
for follow-up observations. Given the early stage of HWO's development, the
design trade space includes almost every aspect of the observatory and
maximizing the yield is a critical task. Even seemingly small changes in
the observatory's design can have a significant and compounding impact on
the mission's yield, as shown by Ref.~\citenum{starkPathsRobust2024}. To
continue that effort, we will focus on \rev{two poorly understood parameters that
affect HWO's scheduling efficiency:} the
observatory's instantaneous field of regard and the number of
characterization observations required to ``characterize'' an EEC.

A primary driver of scheduling efficiency is the FoR. The FoR is the region of
the sky accessible to the observatory at a given time without compromising
thermal stability, power generation, or other spacecraft
constraints\cite{2016jdox,menzel_design_2023}. For space
observatories operating at L2, the FoR is primarily set by the maximum
angles the telescope can safely point towards or away from the
Sun\cite{menzel_design_2023}. Because the FoR determines when
individual target stars are observable, it directly affects how efficiently
the telescope can schedule observations during a large EEC survey. A
telescope with a narrow FoR provides only short observing opportunities for
target stars near the ecliptic, which can reduce the mission's overall
yield of planet detections and characterizations.

An open question for HWO is what observations will be required to
``characterize'' an EEC. Generally,
characterization requires that we understand the planet's orbit, model its
atmosphere, and confirm or rule out the presence of biosignatures. Orbit
determination is expected to require at least 3 well-spaced detections, as
was used by the HabEx and LUVOIR teams and suggested by
Ref.~\citenum{blunt_orbits_2017}. Understanding how many
observations are required to model the planet's atmosphere poses a more
complex challenge and is an area of active research. Recently, a decision
tree observation strategy for HWO was proposed by
Ref.~\citenum{young_retrievals_2024}, suggesting a sequence of 1-4
targeted observations across the UV, visible, and near-infrared
spectrum. This represents a significant unknown for the HWO mission,
because characterization observations can take days to weeks of
dedicated observing time, and currently there have been no studies that quantify
how the additional characterizations will impact the mission's yield.

The interaction between strict FoR constraints and multi-visit characterization
strategies creates a complex, time-dependent scheduling problem. Traditional
yield calculations often optimize integration times for each star without
necessarily carrying out the full scheduling process (e.g. AYO). However, to
accurately assess the scheduling problem at hand, we must simulate the mission
dynamically, enforcing time-domain constraints and modeling the information
gain (e.g. orbital knowledge) as planets are discovered and tracked during the
mission.

In this study, we utilize EXOSIMS to
understand how the FoR and characterization requirements impact the final
yield. We introduce a novel scheduling algorithm to model information gain
during the mission and optimize the transition from discovery to
characterization. We explore a multi-dimensional trade space to quantify HWO's
sensitivity to FoR, characterization requirements, mission duration, and
aperture size. For the purposes of this study, we are defining ``yield'' as the
number of EECs that reach the required number of characterizations ($N_\text{char}$).

The remainder of this paper is structured as follows.
\Cref{sub:field_of_regard} defines the FoR and provides
a table of FoR constraints for relevant space telescopes.
\Cref{sec:orbix_scheduler} introduces our novel dynamic scheduling algorithm.
\Cref{sec:methods_parameters} outlines the study parameters and assumptions we
make for modeling HWO. \Cref{sec:results} presents the results of our yield
calculations and \Cref{sec:discussion} discusses the implications for HWO's
design, limitations, and future work.

\subsection{Field of Regard Definition}
\label{sub:field_of_regard}

\begin{figure}
\centering
\includegraphics[width=0.85\linewidth]{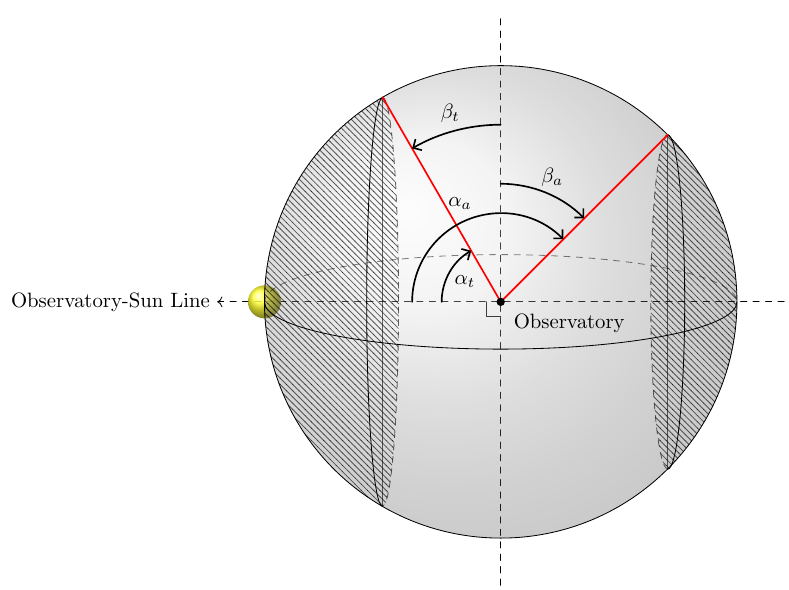} 
\caption{Geometry of an L2 observatory's instantaneous Field of Regard (FoR). The
FoR is defined by the angle between the observatory-Sun line and
the observatory-target line. The accessible region
(unhatched) is constrained by the minimum and maximum angles. Two
conventions are shown: the $\alpha$ angles measure the angle directly from the
Sun line, while the $\beta$ angles measure the deviation from the perpendicular
plane ($90\degree$). The hatched regions represent the solar exclusion zones at
the epoch shown.}
\label{fig:field_of_regard}
\end{figure}

The field of regard defines the region of the sky accessible to the
observatory at any given time. For observatories operating at L2,
the FoR is fundamentally constrained by the need to maintain the spacecraft's
orientation relative to the Sun for thermal stability and power generation. 

The science availability of targets is determined by the angular separation
between the observatory-Sun direction and the observatory-target direction. For
typical L2 observatory designs, rotation maneuvers about the observatory-Sun
line do not significantly alter the orientation of the sunshield relative to
incoming sunlight, allowing for a $360\degree$ azimuthal rotation. Therefore,
the FoR forms an instantaneous annulus on the celestial sphere set by the
limiting angles the telescope can safely point toward and away from the Sun.

There are two primary conventions used to define the limiting Sun angles, both
illustrated in \Cref{fig:field_of_regard}.

\textbf{Convention 1: Sun Angle Reference ($\alpha$)}. This convention, used by
the Roman Space Telescope team \cite{NancyGrace,RomanSpace}, defines the limiting
angles $\alpha$ based on the angle between observatory-Sun line and the
observatory-target line.
\begin{itemize}
    \item $\alpha = 0\degree$: Pointing directly at the Sun.
    \item $\alpha = 180\degree$: Pointing directly anti-Sun.
\end{itemize}
The FoR is the annulus defined by the minimum allowed angle towards the Sun
($\alpha_t$) and the maximum allowed angle away from the Sun ($\alpha_a$).

\textbf{Convention 2: Perpendicular Reference ($\beta$)}. This convention, used
by the JWST team \cite{menzel_design_2023}, defines the limiting angles
relative to the plane perpendicular to the Observatory-Sun line
($\alpha=90\degree$).
\begin{itemize}
    \item $\beta_t$: Maximum angle towards the Sun.
    \item $\beta_a$: Maximum angle away from the Sun.
\end{itemize}

These conventions are related by a $90\degree$ offset: $\beta_t = 90\degree -
\alpha_t$ and $\beta_a = \alpha_a - 90\degree$. The total angular width of the
FoR is $\alpha_a - \alpha_t = \beta_t + \beta_a$. 

To calculate the instantaneous fraction of the sky available to the telescope, we
integrate over the accessible annulus:
\begin{equation}
A =\int_{0}^{2\pi}\int_{\alpha_t}^{\alpha_a} \sin\theta d\theta d\phi=2\pi(\cos \alpha_t - \cos \alpha_a)
\end{equation}
where $\theta$ is the polar angle ($\alpha$) and $\phi$
is the azimuthal angle. Given that the total area of the sky is $4\pi$
steradians, the fraction of the sky instantaneously available is:
\begin{equation}
\frac{\cos \alpha_t - \cos \alpha_a}{2}.
\label{eq:sky_coverage}
\end{equation}
\rev{\Cref{tab:for_summary} lists the FoR constraints for several relevant
space telescopes and mission concepts along with the instantaneous sky coverage of each.}

\begin{table}[t]
\centering
\caption{Field of regard (FoR) constraints for selected space telescopes.
    Instantaneous sky coverage is calculated using \Cref{eq:sky_coverage}.}
\label{tab:for_summary}
\begin{tabular}{lccccc}
\hline
\textbf{Telescope} & $\boldsymbol{\alpha_t}$ & $\boldsymbol{\alpha_a}$ & $\boldsymbol{\beta_t}$ & $\boldsymbol{\beta_a}$ & \textbf{Sky Coverage [\%]} \\
\hline
JWST\cite{menzel_design_2023} & 85\degree & 135\degree & 5\degree & 45\degree & 39.7 \\
Roman\cite{NancyGrace} & 54\degree & 126\degree & 36\degree & 36\degree & 58.8 \\
HabEx\cite{gaudiHabitableExoplanet2020} & 45\degree & 135\degree & 45\degree & 45\degree & 70.7 \\
LUVOIR\cite{TheLUVOIRTeam2019} & 45\degree & 180\degree & 45\degree & 90\degree & 85.4 \\
HWO (nominal)\cite{feinberg_personal_2025} & 45\degree & 135\degree & 45\degree & 45\degree & 70.7 \\
\hline
\end{tabular}
\end{table}

\section{Methods}
\label{sec:orbix_scheduler}

The mission simulations presented in this study utilize a novel, dynamic
scheduling algorithm, implemented in the EXOSIMS framework as a new scheduler
included in the EXOSIMS-Plugins library. This scheduler is designed to optimize
the transition from planet discovery to characterization by incorporating
real-time orbital uncertainty and detection probability calculations. It
employs a greedy heuristic strategy that dynamically balances the discovery of
new EECs (blind search) with the follow-up required for orbit determination and
characterization (targeted search), leveraging the JAX-based \texttt{orbix}
library for high-performance orbital mechanics (in development, see
\Cref{sec:orbix}). \rev{A greedy heuristic selects the locally optimal action
at each decision point, choosing the target with the highest instantaneous
completeness gain rate rather than searching over future decision sequences.
While the blind search target selection is greedy, the scheduler does account
for future constraints by blocking time for scheduled follow-up observations
and reserving time for the expected remaining detections and characterizations
of known planets (see \Cref{sub:orbix_prioritization}).}

\rev{This scheduler differs from the scheduling approach used in previous
EXOSIMS yield studies\cite{morgan19} in several respects. Previous schedulers
used pre-computed dynamic completeness lookup tables to decide when to revisit
a target with no detected exoplanet. Our scheduler removes specific orbits from
the valid set after each non-detection, conditioning the completeness on the
actual observational evidence collected during the mission
(\Cref{subsub:orbix_evidence}). For planets with a detection, previous
schedulers would calculate a wait time before revisiting based on an assumed
orbital period and fractional orbit wait time, returning for follow-up
observations after the wait time is passed. The scheduler we present forecasts
each planet's detection probability at future epochs using its orbital ensemble
and optimizes follow-up timing over a planning horizon
(\Cref{sub:orbix_prioritization}). Our scheduler reserves integration time for
queued characterizations so that new discoveries do not consume time required to
carry out a necessary characterization. 

Additionally, previous EXOSIMS studies often assume that characterization
observations occur at quadrature, placing the planet at a favorable phase angle
regardless of the location of the planet along its actual
orbit\cite{morgan_hwo_2024}. This is done to approximate a more efficient
scheduler without the additional algorithmic complexity or computation cost.
Our scheduler instead observes the planet at its true position along its orbit,
which produces more realistic characterization integration times and scheduling
constraints.}

\subsection{Mathematical Framework and Core Metrics}
\label{sub:orbix_math_framework}

The scheduler's decisions are driven by probabilistic metrics derived from
ensembles of potential orbits. Our framework formally separates the concepts of
an instrument's physical capabilities from the evolving state of knowledge
about a given target.

\subsubsection{Evidence}
\label{subsub:orbix_evidence}
A core feature of the scheduler is how it tracks evidence collected
from each observation. Previous EXOSIMS schedulers pre-calculated a set of
dynamic completeness values for each target star by using the method described
in Ref.~\citenum{brownNewCompleteness2010} and storing the asymptotic dynamic
completeness value. This is a reasonable heuristic, but it assumes that
follow-up observations performed at different times will always provide the
same amount of information. Here, we introduce a more powerful
framework that tracks collected evidence during an EXOSIMS simulation.

We decompose cumulative evidence by scope (star vs. planet) and by type
(detection vs. non-detection). After $k$ observations, our evidence set is given by
\begin{equation}
\label{eq:evidence_decomp}
\mathcal{E}^s_k \;\equiv\; \Big(\, \mathcal{E}_{k}^{\mathrm{s},-},\; \{\,\mathcal{E}_{k}^{\mathrm{p}_i,+}\}_{i} \Big),
\end{equation}
where $\mathcal{E}_{k}^{\mathrm{s},-}$ collects star-level non-detections (blind-search
observations during which a planet meeting sensitivity limits would have been
detected), and $\mathcal{E}_{k}^{\mathrm{p}_i,+}$ collects
detections associated with planet $p_i$. Currently we assume that each detection
of a planet is unique and not confused with another planet.

\subsubsection{Detectability Constraints}
\label{subsub:orbix_detectability}

To quickly determine if an observation of a target planet will reach the
required SNR we utilize the two constraints described by
Ref.~\citenum{brownSingleVisitPhotometric2005}. First, the geometric constraint
states that the planet's apparent separation must be within the instrument's
IWA and OWA. Second, the photometric constraint states that the planet-star
difference in magnitude ($\Delta\text{mag}$) must be less than the instrument's
limiting difference in planet-star magnitude ($\Delta\text{mag}_0$). While the
geometric constraint is simple to compute, $\Delta\text{mag}_0$ is a complex
function that relies on inverting the telescope's exposure time calculator (ETC). For
our work we use \rev{an ETC based on the Roman Coronagraph Instrument's
model\cite{Nemati2014,Nemati2020a,nemati_analytical_2023}, and
invert it using the approach}
described in Ref.~\citenum{spohn_refined_2025}
which parameterizes $\Delta\text{mag}_0$ as a function of integration time
($t_\text{int}$), angular separation ($\alpha$), local zodiacal light ($Z$),
and an exozodiacal light scale factor ($k_{\text{EZ}}$).
\rev{See \Cref{subsub:etc} for more details on the ETC.}

We define the parameter space of an observation as $\mathcal{S} = \{ (o, t,
t_\text{int}) \}$, where $o$ is an orbit, $t$ is the observation start time,
and $t_\text{int}$ is the integration time.  The subset of
detectable states, $D \subseteq \mathcal{S}$, is calculated by comparing
each orbit's apparent separation ($\alpha(o,t)$) and magnitude difference
($\Delta\text{mag}(o,t, t_\text{int})$) to our two constraints:
\begin{equation}
D = \left\{ (o, t, t_\text{int}) \in \mathcal{S} 
    \;\middle|\; \alpha(o,t) \in
    [\text{IWA}, \text{OWA}] 
    \land \Delta\text{mag}(o,t, t_\text{int}) 
    \le \Delta\text{mag}_0(t_\text{int}, \alpha(o,t), Z(t), k_{\text{EZ}}) \right\}.
    \label{eq:detectability_constraint}
\end{equation}
The corresponding indicator function for instrumental detectability,
$\Bbbbone_D: \mathcal{S} \to \{0, 1\}$, is then:
\begin{equation}
\label{eq:detectability}
\Bbbbone_D(o, t, t_\text{int}) = 
\begin{cases}
1 & \text{if } (o, t, t_\text{int}) \in D \\
0 & \text{otherwise}
\end{cases}.
\end{equation}

\subsubsection{Evidence-Conditioned Validity}
\label{subsub:orbix_validity}

For each target star $s$ we randomly generate a large ensemble of orbits
$\mathcal{O}_s$, typically 10,000 orbits in the habitable zone. Define the
evidence-conditioned valid set $V_s(\mathcal{E}^s_k) \subseteq \mathcal{O}_s$ as
the subset of orbits consistent with $\mathcal{E}^s_k$.
\rev{An orbit is ``consistent'' with the accumulated evidence if none of the
prior non-detections of the star would have detected it. Conversely, an orbit
that falls within the detectable set $D$ during a non-detection is ruled out,
because a planet on that orbit would have been detected but was not.}
The corresponding
indicator $\Bbbbone_{V_s(\mathcal{E}^s_k)}: \mathcal{O}_s \to \{0, 1\}$ is
\begin{equation}
\Bbbbone_{V_s(\mathcal{E}^s_k)}(o) =
\begin{cases}
1 & \text{if } o \in V_s(\mathcal{E}^s_k) \\
0 & \text{otherwise}.
\end{cases}
\end{equation}

\rev{After a failed detection observation we update $V_s(\mathcal{E}_k)$ by removing
orbits from $V_s(\mathcal{E}^s_k)$ that would have been detectable:}
\begin{equation}
\label{eq:completeness_update}
V_s(\mathcal{E}^s_{k+1}) = V_s(\mathcal{E}^s_k) \setminus \{o \mid (o, t, t_\text{int}) \in D\}.
\end{equation}
\rev{In other words, any orbit that should have produced a detectable signal
during the observation but did not is removed from the valid set. This binary
filtering is a computationally efficient approximation, a more rigorous
approach would weight orbits by their posterior probability given the evidence
(see \Cref{sub:orbix_uncertainty}).}

\subsubsection{Completeness ($C$)}
\label{subsub:orbix_completeness}

The completeness of a target star $s$, conditioned on the current evidence $\mathcal{E}^s_k$, becomes
\begin{equation}
\label{eq:dynamic_completeness}
C_s(t, t_\text{int} \mid \mathcal{E}^s_k) = \frac{1}{\mathbf{card}(\mathcal{O}_s)} \sum_{o \in \mathcal{O}_s} \Bbbbone_{V_s(\mathcal{E}^s_k)}(o) \cdot \Bbbbone_{D}(o, t, t_\text{int}).
\end{equation}
where $\mathbf{card}(\mathcal{O}_s)$ is the cardinality (number of elements in
the set) of the ensemble of possible orbits for target $s$.
\rev{After each non-detection, $V_s(\mathcal{E}^s_k)$ is updated via
\Cref{eq:completeness_update} and $C_s$ is recalculated.}

\subsubsection{Probability of Detection ($P_\text{det}$)}
\label{subsub:orbix_pdet}

Completeness, as we have formulated it, only informs the search
for new planets. It begins with a large
ensemble of possible orbits and progressively prunes that ensemble based on
non-detections. In contrast, we formulate $P_\text{det}$ to inform observations
once we have positive evidence that a target planet exists by simulating a
localized ensemble of orbits consistent with the detection.

For a known planet $p$ around target star $s$ the probability of
detection conditioned on the evidence is
\begin{equation}
\label{eq:pdet}
P_{\text{det}, p}(t, t_\text{int} \mid \mathcal{E}^s_k) = \frac{1}{\mathbf{card}(\mathcal{O}_p(\mathcal{E}^s_k))} \sum_{o \in \mathcal{O}_p(\mathcal{E}^s_k)} \Bbbbone_{D}(o, t, t_\text{int}),
\end{equation}
where $\mathcal{O}_p(\mathcal{E}^s_k)$ is the localized ensemble of orbits
consistent with $\mathcal{E}^s_k$ (see \Cref{sub:orbix_uncertainty}).

\subsection{Scheduling Strategy and Optimization}
\label{sub:orbix_scheduling}

The scheduler manages a time-sorted queue of observations and 
tries to balance the blind search for new exoplanets and prioritized follow-ups
of known exoplanets. \rev{The FoR constrains this process
primarily by filtering out targets during blind search
(\Cref{eq:blind_search_optimization}) and limiting the observable windows
for follow-up optimization (\Cref{eq:followup_optimization}).}

\subsubsection{The Scheduling Loop}
\label{subsub:orbix_loop}

The scheduling algorithm operates in a continuous loop, maintaining a
time-sorted queue of future observations. Each loop begins by identifying the
target star with the highest instantaneous completeness gain rate ($dC/dt$) for
a potential blind search observation. Before scheduling this observation, the
scheduler verifies feasibility by checking that the target is within the FoR,
that the observation does not conflict with higher-priority queued
observations, and that sufficient total observing time remains in the mission
budget to complete both the new observation and all currently queued
follow-ups. This ensures that new discoveries do not jeopardize the
characterization of known candidates. If these conditions are met, the blind
observation is added to the queue.

The loop then executes the first observation in the queue. When an observation
is executed, the scheduler's response depends on the outcome. A successful
detection or characterization advances the planet's tracking stage (\Cref{subsub:planet_track}), refines its
orbital ensemble (\Cref{sub:orbix_uncertainty}), and immediately adds a follow-up observation to the queue. In
the case of a non-detection during a blind search, the scheduler updates the
target's completeness forecast by removing orbits that should have been
detected and recalculates the target star's completeness (similar to the
``dynamic completeness'' described in Ref.~\citenum{brownNewCompleteness2010}).
If a follow-up observation of a known planet fails, future observations are
given more integration time on the assumption that the modeled parameters
(e.g., geometric albedo) are overestimating the true planet's brightness.
Because integration times are chosen from \rev{a fixed grid of 50
logarithmically spaced values between 1 hour and 60 days}, we apply a simple
heuristic to increase the integration time: after $n_\text{failures}$ in a
row, we search up to $2n_\text{failures}$ array steps toward longer integration
times and select the longest integration time that still yields a valid observation. If
three consecutive characterization attempts fail, the planet is retired.

\subsubsection{Blind Search Optimization}
\label{subsub:orbix_blind}

The objective during the discovery phase is to maximize the rate of
completeness gain. \rev{For an observation beginning at time $t$, the}
figure of merit \rev{is} the completeness normalized by the integration time:
\begin{equation}
J_{\text{blind}}(s, t, t_\text{int}) = \frac{C_s(t, t_\text{int})}{t_\text{int}}.
\end{equation}
To enable rapid computation, \rev{we pre-forecast $J_\text{blind}$ by evaluating
it over a two-dimensional grid of 100 evenly spaced candidate observation
epochs $t$ spanning the remaining mission lifetime and the 50 log-spaced
integration times $t_\text{int}$ defined above. At each epoch, the optimal integration time}
$t_{\text{int},s}^*(t) = \arg\max_{t_{\text{int}}} J_{\text{blind}}(s, t,
t_{\text{int}})$ and the corresponding maximum rate $J^*(s, t)$ are
\rev{cached} and dynamically updated for each target star $s$. \rev{When the
scheduler must select its next target at the current simulation time $t_c$, it
looks up the pre-forecasted values rather than recomputing from scratch:}
\begin{equation}
\label{eq:blind_search_optimization}
s^* = \arg\max_{s \in S_{\text{avail}}(t_c)} J^*(s, t_c),
\end{equation}
where $S_{\text{avail}}$ is the set of stars satisfying FoR constraints and
other filters, such as the maximum number of visits to the target star.

\subsubsection{\rev{Planet Tracks}}
\label{subsub:planet_track}

\rev{A detection occurs when a simulated observation achieves an SNR exceeding the
broadband detection threshold (SNR $\ge$ 7 in this study) at the planet's true position.
Upon detection, a planet track $p$ is initiated and the planet immediately
enters the follow-up queue. The ``track'' represents the planet's fixed
observational pathway from initial detection to full characterization
through a sequence of observational stages.

The stage number equals the planet's count of successful observations: stage~1
after the initial detection, stage~2 after the first successful follow-up, and
so on. A planet must complete $N_\text{det}$ detection stages before becoming
eligible for spectral characterization, then $N_\text{char}$ characterization
stages to be considered fully characterized. The stage governs both the
planet's scheduling priority (\Cref{sub:orbix_prioritization}) and the orbital
uncertainty assigned to its ensemble (\Cref{sub:orbix_uncertainty}).}

\subsubsection{Follow-up Optimization}
\label{subsub:orbix_followup}

\rev{Follow-up observations are prioritized over blind searches according to
the planet's track stage (\Cref{sub:orbix_prioritization}), not by inferred
planet properties. We assume all detected EECs are of equal scientific
interest.} The scheduler optimizes the timing $t^*$ and duration
$t_{\text{int}}^*$ within a planning horizon $H$ (e.g., 250 days) to maximize
$P_{\text{det}}$ efficiency, incorporating a small penalty for later times to
favor prompt follow-up:

\begin{equation}
\label{eq:followup_optimization}
(t^*, t_\text{int}^*) = \arg\max_{t, t_\text{int}} \left( \frac{P_{\text{det}, p}(t, t_\text{int})}{t_\text{int}} - \lambda (t - t_c) \right),
\end{equation}
where $\lambda$ is a small factor ($10^{-8}$) such that when $P_{\text{det},
p}(t, t_\text{int})$ is the same between multiple observations, the scheduler selects
the observation closest to the current time. This optimization is subject to
constraints:
\begin{enumerate}
    \item $P_{\text{det}, p}(t, t_\text{int}) \ge P_{\text{thresh}}$ (e.g., 95\%, relaxed if necessary).
    \item $t \in [t_c + t_{\text{wait}}, t_c + H]$, where $t_{\text{wait}}$ is the minimum revisit time (e.g., 15 days).
    \item Observability (FoR) during the observation window $[t, t+t_\text{int}+t_\text{OH}]$, where $t_\text{OH}$ is the overhead time.
    \item Availability in the schedule queue.
\end{enumerate}

\subsection{Orbital Uncertainty Management}
\label{sub:orbix_uncertainty}

A fundamental and novel component of the scheduler is the dynamic
management of orbital uncertainty. The uncertainty of the ensemble
$\mathcal{O}_p(t)$ is systematically reduced with each successful follow-up
observation, \rev{advancing the planet's stage
(\Cref{subsub:planet_track}).}

Currently, this system uses an idealized ensemble-based approximation that is
consistent with the probabilistic framework used elsewhere in this work. Upon
the first detection, we instantiate an ensemble of \rev{$N=500$} possible
orbits\rev{, chosen so that the binomial standard error of $P_\text{det}$
remains below $\sim1\%$ at the 95\% scheduling threshold
($\sqrt{p(1-p)/N}\approx0.010$ at $p=0.95$),}
\begin{equation}
\mathcal{O}_p = \{\,\boldsymbol{\theta}^{(k)}\,\}_{k=1}^{500},\quad \boldsymbol{\theta}=(a,e,i,\Omega,\omega,M_0, A_g)\,,
\end{equation}
by randomly sampling parameters around the simulated truth. \rev{Here $a$ is the
semi-major axis, $e$ the eccentricity, $i$ the inclination, $\Omega$ the
longitude of the ascending node, $\omega$ the argument of periapsis, $M_0$ the
mean anomaly at epoch, and $A_g$ the geometric albedo.} \rev{We draw $a$ and
$A_g$ from log-normal distributions centered on the true values, while the
remaining elements are held fixed at their true (simulated) values. A
log-normal distribution was chosen because they
are strictly positive quantities and the log-normal ensures that sampled values
remain physical. In log-space, the distribution is symmetric and the standard
deviation $f$ has a straightforward interpretation as a fractional uncertainty.}
We selected $a$ and $A_g$ to ensure that
$\mathcal{O}_p$ has uncertainty in both $\alpha$ and $\Delta\text{mag}$, corresponding
to both constraints in \Cref{eq:detectability_constraint}, while
maintaining a computationally feasible ensemble size.

For any candidate follow-up epoch $t$, each $\boldsymbol{\theta}^{(k)}$ is
propagated to $t$ to produce predicted apparent separation and flux-ratio
quantities that feed the Monte Carlo estimate of $P_{\text{det},p}(t,
t_{\text{int}})$. With each successful follow-up, the scheduler advances the
planet's stage, lowers $f$ (20\% on initial detection, then 10\% on second,
and 2.5\% on third), and generates a new ensemble $\mathcal{O}_p$. \rev{We
chose these values to be deliberately optimistic. Prior work on orbital
parameter retrieval, such as the simulations presented in the HabEx Final
Report\cite{gaudiHabitableExoplanet2020}, suggests that a small number of
well-spaced direct imaging detections can constrain the semi-major axis and
eccentricity of an exo-Earth to better than 10\% ($1\sigma$). However, those
results assume well-spaced observations, while our scheduler prioritizes
high-$P_\text{det}$ follow-up windows and does not yet optimize for orbital phase
coverage. How orbital uncertainty evolves as a function of observation timing,
phase coverage, and orbital geometry across the full
parameter space relevant to yield modeling remains an open
problem. Rather than implement a more complex
but still unvalidated heuristic, we adopted a simple, favorable model that
allows us to isolate the scheduling effects of the FoR and $N_\text{char}$
without conflating them with the performance of any particular orbit-fitting
approach.}

The idealized shrinkage rule
provides a computationally light alternative to full orbit fitting, and future
development will add the option to replace it with
orbit fitting in a Bayesian framework, updating the $\mathcal{O}_p$ to
approximate the posterior $p(\boldsymbol{\theta}\mid \mathcal{E}^s_k) \propto
\mathcal{L}(\mathcal{E}^s_k\mid \boldsymbol{\theta})\,p(\boldsymbol{\theta})$
from the collected evidence $\mathcal{E}^s_k$, rather than perturbing around
the true (simulated) values. \rev{Incorporating orbit fitting after each detection
and characterization observation will result in at least an order-of-magnitude
increase in computational cost per simulation, making a full exploration of the
trade space studied here prohibitive. We note that incorporating realistic orbital
uncertainty would likely narrow the scheduling windows in which
$P_\text{det}$ exceeds the observation threshold for some planets,
disproportionately penalizing narrow-FoR configurations that already
have fewer scheduling opportunities.
Quantifying this interaction is a goal of our future work.}

\subsection{Prioritization and Time Budget Constraints}
\label{sub:orbix_prioritization}

\rev{Prioritization is governed by the planet's track stage
(\Cref{subsub:planet_track}). Higher-stage observations (e.g.,
characterizations or final orbit-determination detections) receive higher
priority.} If a high-priority follow-up conflicts with a scheduled
lower-priority observation, the lower-priority observation is ``bumped'' and
rescheduled.

The scheduler checks each observation to ensure that the scheduler is not
scheduling observations that would exceed the total survey time budget. A new
blind search of duration $T_{\text{new}}$ (including overheads) is only
scheduled if sufficient time remains:
\begin{equation}
T_{\text{spent}} + T_{\text{queued}} + T_{\text{reserved}} + T_{\text{new}} \leq T_{\text{survey}}.
\end{equation}
Here, $T_{\text{spent}}$ is the time used, $T_{\text{queued}}$ is the time for
actions currently in the queue, and $T_{\text{reserved}}$ is the estimated time
required to achieve the required $N_{\text{det}}$ and $N_{\text{char}}$ for all
active planet tracks. This reservation mechanism ensures that discovery efforts
do not jeopardize the characterization of existing candidates. 

To compute $T_{\text{reserved}}$ we sum
the time required to complete the remaining detections and
characterizations for each (non-deferred) track using per-mode estimates of integration time
and the mode overheads (slew/settling plus instrument overhead):
\begin{equation}
\label{eq:t_reserved}
T_{\text{reserved}}\;=\;\sum_{\text{tracks}}
\Big[\,N^{\text{rem}}_{\text{det}}\,\big(t^{\ast}_{\text{det}}+t^{\text{oh}}_{\text{det}}\big)
\;+\;N^{\text{rem}}_{\text{char}}\,\big(t^{\ast}_{\text{char}}+t^{\text{oh}}_{\text{char}}\big)\,\Big].
\end{equation}
Here $N^{\text{rem}}_{\text{det}}$ and $N^{\text{rem}}_{\text{char}}$ are the
remaining detections and characterizations needed for the track to reach full
characterization. The terms $t^{\text{oh}}_{\text{det}}$ and
$t^{\text{oh}}_{\text{char}}$ are the per-observation overheads for detection
and characterization modes, respectively. The exposure-time estimates
$t^{\ast}_{\text{det}}$ and $t^{\ast}_{\text{char}}$ are set using
a heuristic value of 15 days of integration time, or 25\% of the maximum allowed
time for a single observation. In testing we found that this heuristic avoids
over-reserving time, which suppresses the discovery of new planets, while still
ensuring that sufficient time is budgeted for characterization.


\section{Study Parameters and Assumptions}
\label{sec:methods_parameters}

To evaluate the impact of the FoR, characterization requirements, telescope
aperture, and mission duration on the scientific yield of HWO, we utilized
EXOSIMS~\cite{delacroix_science_2016}. The simulations employ the
dynamic scheduling methodology described in \Cref{sec:orbix_scheduler} as
implemented in the \texttt{OrbixScheduler}, \texttt{OrbixCompleteness}, and
\texttt{OrbixUniverse} modules. This section details the specific
parameters, models, and assumptions defining the study's trade space and fixed
constraints.

\subsection{Study Trade Space}

This study explores a multi-dimensional parameter space, summarized in
Table~\ref{tab:trade_space}. We analyze the sensitivity of the mission yield
across different mirror diameters, dedicated EEC survey time, characterization
requirements, and FoR constraints. For every unique combination of these
parameters, 100 EXOSIMS simulations were executed to ensure statistical
robustness.

\begin{table*}[ht!]
    \centering
    \caption{Explored trade space parameters}
    \label{tab:trade_space}
    \begin{tabular}{lll}
        \toprule
        \textbf{Parameter} & \textbf{Symbol} & \textbf{Values Explored} \\
        \midrule
        Inscribed aperture diameter & $D_\text{insc}$ & [6.5, 8.0] m \\
        Required characterizations & $N_\text{char}$ & [1, 2, 3, 4] spectra per planet \\
        Dedicated survey time & $T_\text{survey}$ & [2.5, 5.0, 7.5] years \\
        Field of regard (FoR) & ($\beta_t$, $\beta_a$) & 9 paired configurations (see details below) \\
        \bottomrule
    \end{tabular}
\end{table*}

\subsubsection{Aperture}

We investigated two inscribed aperture diameters, 6.5\,m and 8.0\,m. A fixed
obscuration factor of 17.48\% is applied, resulting in physical (circumscribed)
pupil diameters of approximately 7.88\,m and 9.70\,m, respectively. 

\subsubsection{Mission Duration}

We analyzed three allocations of dedicated EEC survey time
($T_\text{survey}$): 2.5, 5.0, and 7.5 years. In our simulations the time not
dedicated to the exoplanet survey is assumed to be utilized by general
astrophysics observations. However, the scheduling algorithm currently assumes
that the EEC survey has priority over all other observations and can use all of
the mission's allocated observing time.

\subsubsection{Characterization Requirements}

A planet is considered ``characterized'' upon achieving the required number of
successful spectral observations ($N_\text{char}$). We varied this requirement
from 1 to 4 spectra per planet based on the decision tree framework in
Ref~\citenum{young_retrievals_2024}. This requirement is subsequent to the 3
successful imaging detections ($N_\text{det}=3$) needed for initial orbit
estimation (see Table~\ref{tab:strategy_params}). For simplicity, we assume
that all characterizations are conducted at the same wavelength, 910\,nm.

\subsubsection{Field of Regard}

We parameterized the FoR by the limiting angles toward the Sun ($\beta_t$) and
away from the Sun ($\beta_a$) (see \Cref{fig:field_of_regard}). We explored
nine paired combinations, incrementally increasing the accessible sky area:
($5\degree$, $10\degree$), ($10\degree$, $20\degree$), ($15\degree$,
$30\degree$), ($20\degree$, $40\degree$), ($25\degree$, $50\degree$),
($30\degree$, $60\degree$), ($35\degree$, $70\degree$), ($40\degree$,
$80\degree$), and ($45\degree$, $90\degree$). These correspond to a total
instantaneous FoR ranging from $15\degree$ to $135\degree$ and sky coverage
between $13\%$ and $85\%$. Note that while it's common for telescopes to design for
$\beta_t = \beta_a$, Ref.~\citenum{spohnHowHabitable2024a} found that yield was driven
by the total FoR angle (i.e., $\beta_t + \beta_a$) and was insensitive to the
ratio between the two limiting angles.

\rev{In practice the FoR is not an independently adjustable parameter. It is
coupled to sunshield geometry, thermal stability requirements, and achievable
wavefront stability, all of which directly affect coronagraphic contrast
performance. Widening the FoR may require engineering trade-offs in these
areas and a complete design optimization would need to consider these
constraints jointly.}

\subsection{Fixed Assumptions and Models}

\subsubsection{Telescope and Instrument Configuration}
\label{subsub:observatory_instrument}

The observatory and instrument suite parameters are based on the preliminary
HWO design named ``Exploratory Analytic Case 1'' (EAC1). The parameters for
EAC1 were created by the Science Engineering Interface working group as part of
the START/TAG program. The parameters are available on the Science Engineering
Interface group's GitHub repository
(\url{https://github.com/HWO-GOMAP-Working-Groups/Sci-Eng-Interface}). Except
for a few study-specific choices, we adopt the EAC1 parameters. All fixed
assumptions and parameter values are summarized in
Table~\ref{tab:instrument_params}.

\begin{figure}[ht!]
    \centering
    \includegraphics[width=1\linewidth]{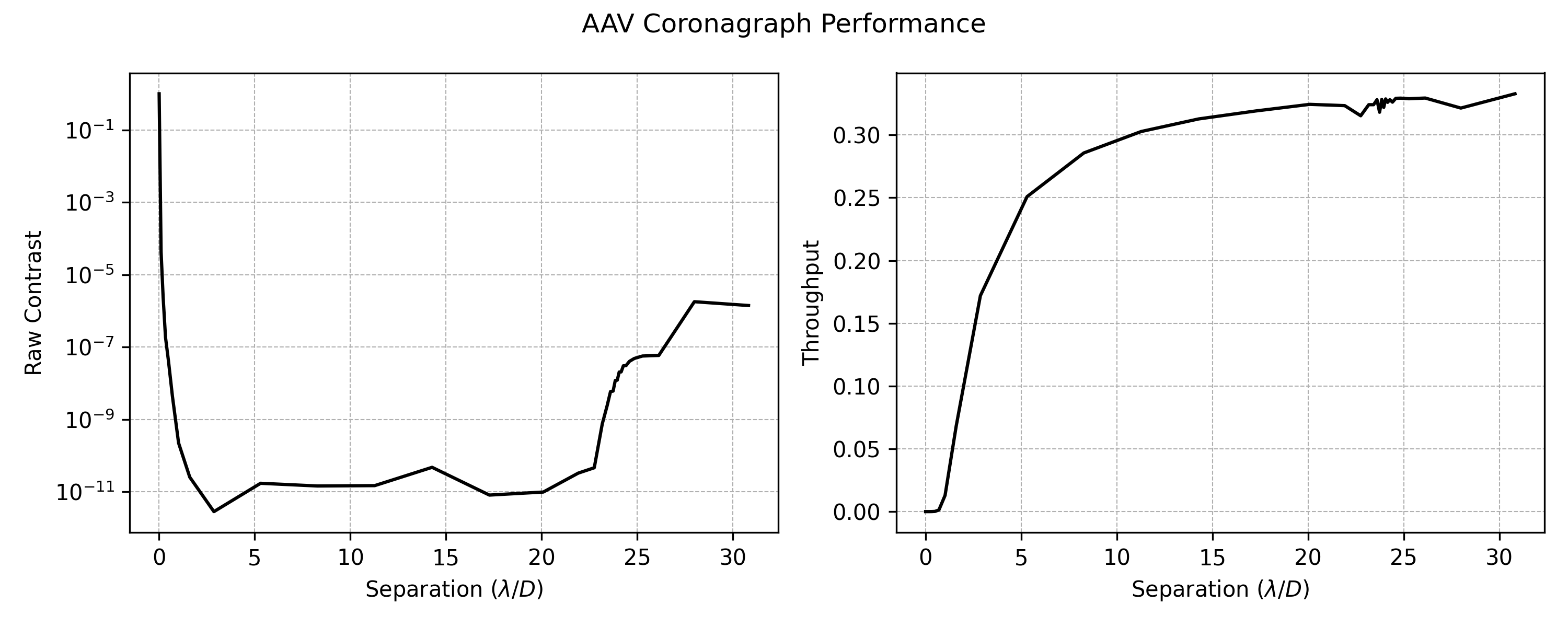}
    \caption{Performance of the charge 4 amplitude apodized vortex coronagraph used in this
    study\cite{mawet_apodized_2025}. The left panel shows the raw contrast and
    the right panel shows the core throughput as a function of angular
    separation in units of $\lambda/D$.}
    \label{fig:coronagraph_performance}
\end{figure}

The coronagraph used in this study is a charge 4 amplitude apodized vortex
(AAV) coronagraph\cite{mawet_apodized_2025} which was created by Susan Redmond,
Dimitri Mawet, and Arielle Bertrou-Cantou as part of the Coronagraph Design
Survey\cite{belikov_coronagraph_2024}. The coronagraph was provided as a
``yield input package'' (YIP), a standardized set of FITS files that yield
calculators use to compute coronagraph performance metrics.
The performance of this coronagraph is shown in \Cref{fig:coronagraph_performance}.

The simulations employ two distinct observing modes:
\begin{enumerate}
    \item \textbf{Detection:} Broadband visible imaging centered at 500\,nm (20\% bandwidth).
    \item \textbf{Characterization:} Lenslet Integral Field Spectrograph (IFS) centered at 910\,nm with $R=140$.
\end{enumerate}
Key telescope and instrument parameters are summarized in
Table~\ref{tab:instrument_params}.

\subsubsection{\rev{Exposure Time Calculator}}
\label{subsub:etc}

The optical system is modeled with EXOSIMS's \texttt{Nemati} module, which is
based on the model adopted by the Nancy Grace Roman Space Telescope's
Coronagraph Instrument
team\cite{Nemati2014,Nemati2020a,nemati_analytical_2023}. The
\rev{\texttt{Nemati} module provides the ETC used to determine integration
times for both detection and characterization and to compute the photometric
constraint $\Delta\text{mag}_0$ (\Cref{subsub:orbix_detectability}). The ETC
computes the signal-to-noise ratio by modeling the planet signal against
background noise sources (such as zodiacal light and exozodiacal light) and a
residual speckle term that acts as a systematic noise floor. The noise floor is
calculated as the raw speckle count rate reduced by a post-processing factor
(Table~\ref{tab:instrument_params}). The \texttt{Nemati} model also provides a
stability factor that can model contrast degradation from temporal wavefront
instability\cite{Nemati2020a}, but we leave this at its default value of unity
because HWO's stability budget remains too preliminary to warrant a specific
value. Because the background noise terms depend on the
planet signal in this model, the inversion described in
Ref.~\citenum{spohn_refined_2025} solves the coupled equation analytically to
produce self-consistent $\Delta\text{mag}_0$ values.} The detector is assumed
to be noiseless.

\rev{For characterization observations the IFS mode collects photons in a spectral channel of width
$\lambda/R$, which is considerably narrower than the broadband detection
filter. Additionally, for characterization the planet's own
photon signal is included in the background noise budget. Together, these
effects make characterization integration times considerably longer than
detection integration times at the same angular separation.}

\begin{table*}[ht!]
    \centering
    \caption{Telescope and Instrument Parameters}
    \label{tab:instrument_params}
    \begin{tabular}{ll}
        \toprule
        \textbf{Parameter} & \textbf{Value} \\
        \midrule
        \multicolumn{2}{l}{\textbf{Telescope and Overheads}} \\
        Obscuration factor & 17.48\% \\
        Slew/settling time & 1 hr \\
        Coronagraph overhead & 2.4 hr \\
        \midrule
        \multicolumn{2}{l}{\textbf{Coronagraph (AAVC)}} \\
        Inner working angle (IWA) & 2.77\,$\lambda/D$ \\
        Outer working angle (OWA) & 32.0\,$\lambda/D$ \\
        System design wavelength & 1000\,nm \\
        Instrument optics transmission & 95\% \\
        Core throughput & See \Cref{fig:coronagraph_performance} \\
        Raw contrast & See \Cref{fig:coronagraph_performance} \\
        Core photometric aperture area & 1.539\,$(\lambda/D)^2$ \\
        \midrule
        \multicolumn{2}{l}{\textbf{Detection Mode (Imaging)}} \\
        Wavelength & 500\,nm \\
        Fractional bandwidth & 20\% \\
        Pixel scale & 4.85\,mas/pixel \\
        Pixel size & 13\,$\mu$m \\
        Instrument optics transmission & 36.2\% \\
        Frame exposure time & 100\,s \\
        Required SNR & 7 \\
        \midrule
        \multicolumn{2}{l}{\textbf{Characterization Mode (IFS)}} \\
        Wavelength & 910\,nm \\
        Spectral resolution (R) & 140 \\
        Pixel scale & 4.85\,mas/pixel \\
        Pixel size & 12\,$\mu$m \\
        Lenslet sampling & 3.5 px/lenslet \\
        Instrument optics transmission & 30.7\% \\
        Frame exposure time & 300\,s \\
        Required SNR & 5 \\
        \midrule
        \multicolumn{2}{l}{\textbf{Common Parameters}} \\
        Read noise & 0 e$^-$\,/pix/read \\
        Dark current & 0 e$^-$\,/pix/s \\
        Clock-induced charge (CIC) & 0 e$^-$\,/pix/frame \\
        Photon-counting efficiency (PCeff) & 0.99 \\
        Quantum efficiency (QE) & 90\% \\
        Post-processing factor (PPF) & 30 \\
        \bottomrule
    \end{tabular}
\end{table*}

\subsubsection{Mission Design and Observation Strategy}

The new EXOSIMS module, \texttt{OrbixScheduler}, manages observations
dynamically. Integration times are calculated to meet the required SNR, but for
performance reasons we restrict them to a grid of 50 logarithmically spaced
times between 1 hour and 60 days. The scheduler uses the in-loop
$P_\text{det}$ calculation to determine when to attempt follow-up observations.
An observation is only scheduled if the $P_\text{det}$ exceeds a threshold of
95\%. A minimum wait time of 15 days is enforced between revisits to the same
target, and the scheduler looks ahead a maximum of 250 days to plan follow-ups
(note that this constraint may be relaxed if no observation meets all
constraints). Only a maximum of 25 visits per star are permitted.

Mission design and strategy parameters are summarized in Table~\ref{tab:strategy_params}.

\begin{table*}[ht!]
    \centering
    \caption{Mission design and observation strategy parameters}
    \label{tab:strategy_params}
    \begin{tabular}{ll}
        \toprule
        \textbf{Parameter} & \textbf{Value} \\
        \midrule
        Observatory location & L2 halo orbit \\
        Required detections ($N_\text{det}$) & 3 \\
        Initial $P_\text{det}$ scheduling threshold & 0.95 (95\%) \\
        Initial follow-up scheduling horizon & 250 days \\
        Minimum revisit wait time & 15 days \\
        Minimum integration time & 1 hour \\
        Maximum integration time & 60 days \\
        Number of discrete integration times (log spaced) & 50 \\
        Maximum visits per star & 25 \\
        Number of failed detections before removal & 10 \\
        Minimum completeness (for target list filtering) & 0.01 \\
        Star V magnitude range & [2, 15] \\
        \bottomrule
    \end{tabular}
\end{table*}

\subsubsection{Astrophysical Assumptions}

The simulated universe is generated based on the star properties from the
EXOCAT1 star catalog\cite{turnbullExoCat1Nearby2015}. The planet population is
generated from the ``nominal'' occurrence rates calculated in
Ref.~\citenum{dulz_joint_2020}.
For all stars the exozodiacal dust is fixed at 3 times the
brightness of the solar system's zodiacal cloud (i.e., $n_\text{EZ}=3$), based
on the results from the Large Binocular Telescope Interferometer HOSTS
survey\cite{ertelHOSTSSurvey2020}. Astrophysical parameters are summarized in
Table~\ref{tab:astro_params}.

\begin{table*}[t!]
    \centering
    \caption{Astrophysical assumptions}
    \label{tab:astro_params}
    \begin{tabular}{ll}
        \toprule
        \textbf{Parameter} & \textbf{Value/Model} \\
        \midrule
        ExoEarth occurrence rate ($\eta_\oplus$) & 0.24 \\
        Eccentricity ($e$) & 0 (Circular orbits) \\
        Phase function ($\Phi$) & Lambertian \\
        Geometric albedo ($A_G$) & 0.2 \\
        Exozodiacal light level ($n_\text{EZ}$) & 3 zodi \\
        V band surface brightness of exozodi ($n_\text{EZ}=1$) & 22 mag/arcsec$^2$ \\
        V band surface brightness of local zodi & 23 mag/arcsec$^2$ \\
        \midrule
        \multicolumn{2}{l}{\textbf{Earth-like planet definition (adopted from HabEx and LUVOIR final reports\cite{gaudiHabitableExoplanet2020,TheLUVOIRTeam2019})}} \\
        Scaled semi-major axis ($a_{s}$)$^a$ & $0.95 \leq a_{s} < 1.67$ AU \\
        Radius ($R_p$) & $(0.8 / \sqrt{a_{s}}) \leq R_p < 1.4$ $R_{\oplus}$ \\
        \bottomrule
    \end{tabular}
    \vspace{1mm}
    \footnotesize
    \\ $^a$ Scaled semi-major axis is defined as $a_{s} = a / \sqrt{L_{star}/L_{\odot}}$.
\end{table*}

\section{Results}
\label{sec:results}

We simulated each configuration on the same set of 100 randomly generated
universes to quantify the sensitivity of HWO's EEC yield across our
multi-dimensional trade space of FoR, required number of characterizations
($N_\text{char}$), dedicated survey time ($T_\text{survey}$), and inscribed
aperture diameter ($D_\text{insc}$). \rev{An example mission schedule and
scheduling constraint visualization are provided in \Cref{sec:example_schedule}.}

\subsection{Sensitivity to the Field of Regard}

\begin{figure}[t]
    \centering
    \includegraphics[width=0.85\linewidth]{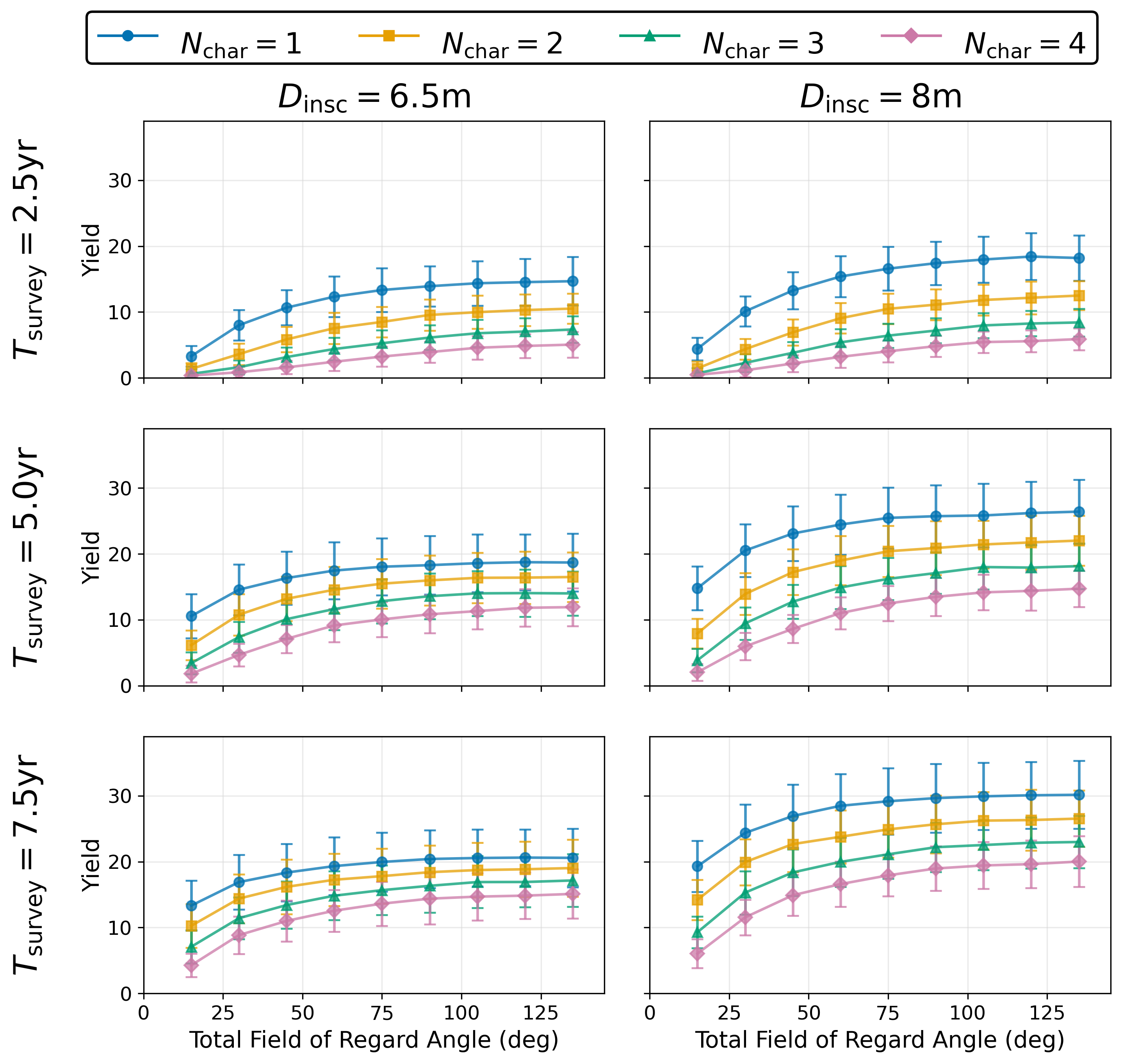}
    \caption{EEC yield as a function of total FoR angle.
    Rows represent dedicated survey times (2.5, 5.0, 7.5
    years) and columns represent inscribed diameters (6.5\,m, 8.0\,m). Colored
    lines indicate the required number of characterizations
    ($N_\text{char}=1\dots4$). Error bars show the 1$\sigma$ scatter across
    100 realizations.}
    \label{fig:unified_obs_diam_nchar}
\end{figure}

\begin{figure}[ht]
    \centering
    \includegraphics[width=0.85\linewidth]{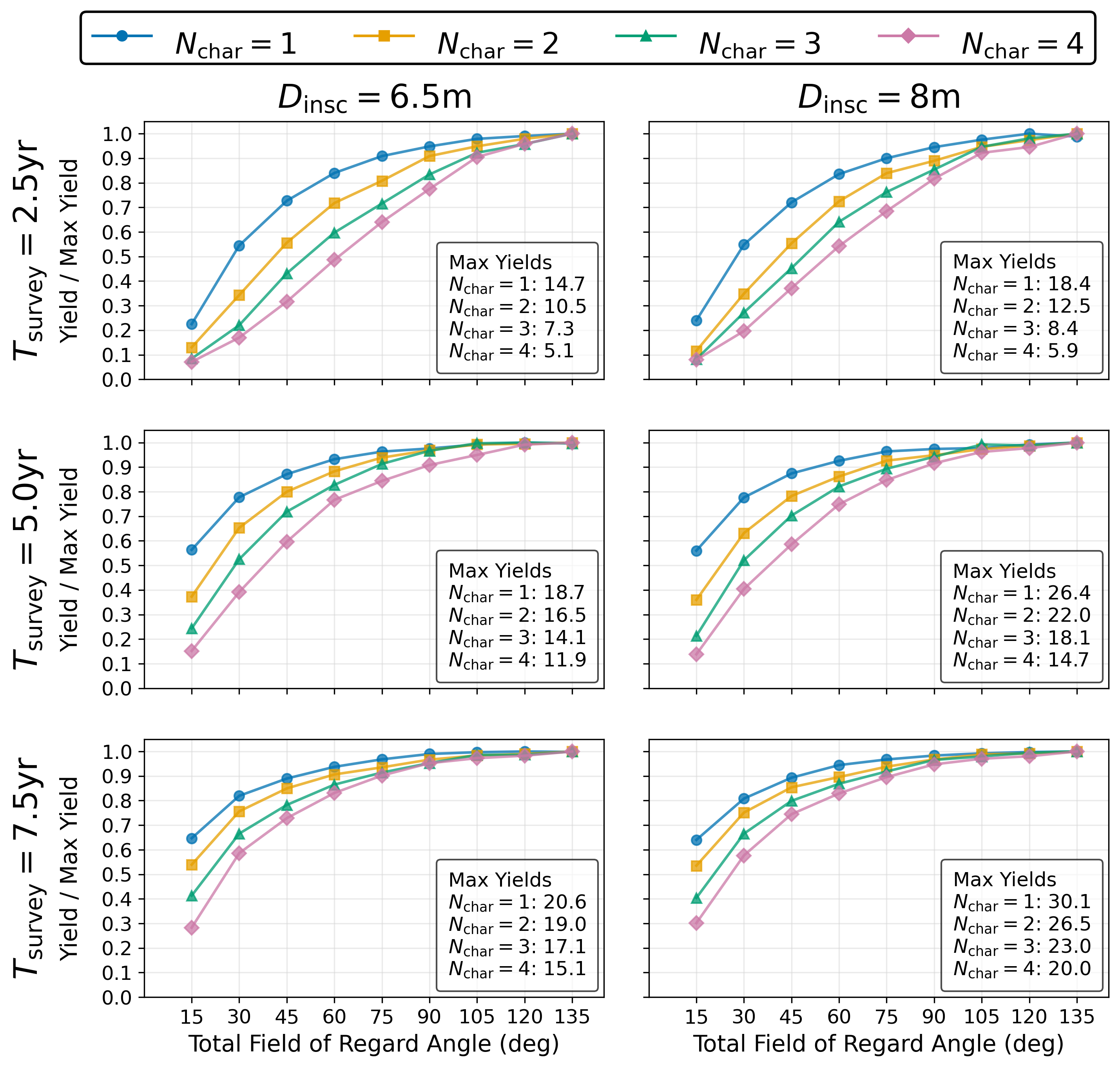}
    \caption{Mean fractional EEC yield normalized by the maximum yield of each configuration
    (e.g. the $N_\text{char}$, $T_\text{survey}$, and $D_\text{insc}$ values)
    as a function of total FoR angle. Rows indicate dedicated survey time (2.5,
    5.0, 7.5 years) and columns indicate inscribed aperture diameter (6.5\,m,
    8.0\,m). Colored lines denote the required number of characterizations
    ($N_\text{char}=1\dots4$).}
    \label{fig:unified_obs_diam_nchar_fractional}
\end{figure}

The absolute yield trends are shown in \Cref{fig:unified_obs_diam_nchar}
and the fractional yield trends are shown in \Cref{fig:unified_obs_diam_nchar_fractional}.
The instantaneous FoR is a major driver of mission yield across all scenarios.
The relationship between FoR and yield is non-linear, with a steep
initial increase followed by pronounced diminishing returns. Expanding the FoR
from $15\degree$ to $60\degree$ yields an average increase of $\sim128\%$.
Increasing from $60\degree$ to $90\degree$ provides a much smaller average gain
of $\sim12\%$, and the curve flattens significantly beyond $90\degree$, where
$90\degree\rightarrow135\degree$ adds only $\sim5.5\%$ on average. The necessity of
a wide FoR is magnified when characterization requirements are increased.
For $N_\text{char}=1$, the gain from $15\degree$ to $75\degree$ is $\sim87\%$, but
for $N_\text{char}=4$, the gain is $\sim310\%$.

\Cref{fig:unified_obs_diam_nchar_fractional} emphasizes the common FoR trend
across mission durations and apertures and highlights the diminishing returns
past roughly $90\degree$. The normalization also makes the relationship between FoR and 
$N_\text{char}$ more directly comparable, demonstrating that higher $N_\text{char}$
sees a greater fractional drop in yield as the FoR is decreased.

\subsection{Impact of Characterization Requirements}


\begin{figure}[ht]
    \centering
    \includegraphics[width=0.90\linewidth]{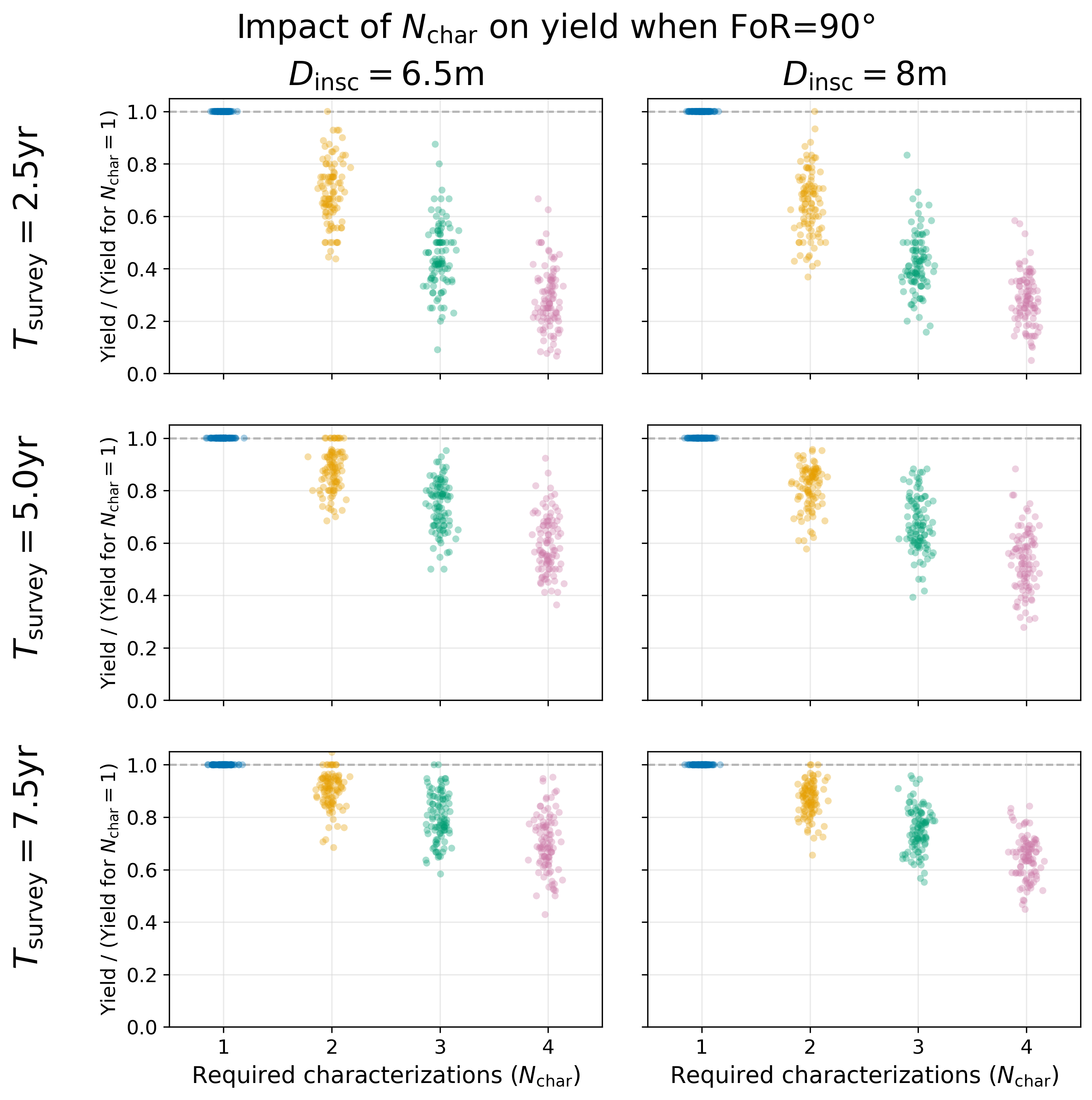}
    \caption{Fractional EEC yield at fixed FoR of $90\degree$ as a function of the
    required number of characterizations ($N_\text{char}$). Values are normalized by
    each universe's yield at $N_\text{char}=1$ for the same $(T_\text{survey}, D_\text{insc})$.
    Rows indicate dedicated survey time (2.5, 5.0, 7.5 yr) and columns indicate
    inscribed diameter (6.5\,m, 8.0\,m). For each panel and $N_\text{char}$ value,
    each point represents the yield of a single randomly generated universe,
    100 in total. Note that each mission design was tested on the same 100 randomly
    drawn universes. The y value of each point is ratio of the yield
    when $N_\text{char}=k$ to the yield when $N_\text{char}=1$, holding
    the simulated universe and all other parameters fixed.}
    \label{fig:slice_for90_nchar}
\end{figure}

Increasing the required number of characterizations imposes a substantial
penalty on total EEC yield, as the time required to characterize each
EEC increases.

\Cref{fig:slice_for90_nchar} compares yield when varying only $N_\text{char}$.
The figure shows all 100 simulations aggregated in
\Cref{fig:unified_obs_diam_nchar} to generate the points for a FoR of
$90\degree$. Each point represents the yield of the simulation at
$N_\text{char}=k$ when compared to the yield of the simulation
$N_\text{char}=1$ with all other parameters fixed (including the generated
planets). We see a clear decrease in fractional yield when increasing
$N_\text{char}$ across all apertures and survey durations. Longer surveys
partially mitigate the loss while the dependence on aperture is weaker.

Averaged across FoR, survey durations, and apertures, the transition from
$N_\text{char}=1\rightarrow2$ results in an absolute loss of $\sim4.1$
EECs, a relative loss of $21.6\%$. Subsequent steps incur similar fractional
penalties: $2\rightarrow3$ reduces yield by $21.5\%$ and $3\rightarrow4$ by
$21.8\%$. Cumulatively, requiring four characterizations instead of one
reduces the mean yield by $51.9\%$.

\subsection{Influence of Survey Duration and Aperture Size}

Survey duration and aperture are dominant determinants of absolute yield.
Additionally, longer surveys can substantially mitigate the characterization
penalty. The relative yield loss for $1\rightarrow2$ characterizations
decreases with survey time:
\begin{itemize}
    \item $T_\text{survey}=2.5$ yr: $\sim38\%$ loss
    \item $T_\text{survey}=5.0$ yr: $\sim20\%$ loss
    \item $T_\text{survey}=7.5$ yr: $\sim14\%$ loss
\end{itemize}
Tripling the survey time from 2.5 to 7.5 years increases yield by $1.76\times$
for $N_\text{char}=1$, but by $4.26\times$ for $N_\text{char}=4$,
reflecting the additional flexibility needed for multi-visit strategies.

As expected, increasing $D_\text{insc}$ increases the total yield. We found
that changing $D_\text{insc}$ from 6.5\,m to 8.0\,m yields an average relative
increase of $1.33\times$ across the trade space and showed no significant
interaction with the other parameters. The larger aperture boosts yield by
reducing integration times and raising the discovery rate. This produces a
larger pool of candidates to characterize, whose follow-up rate remains limited
by the same scheduling constraints as a smaller aperture with its smaller
pool of targets but longer integration times.

\section{Discussion}
\label{sec:discussion}

Rather than prescribing a single path to reaching HWO's EEC target, our
results indicate robust trends and compensating trades between field of regard,
characterization requirements, survey duration, and aperture. 

\begin{itemize}
    \item \textbf{Field of Regard (FoR):} If constraints
    require a FoR below $\sim90\degree$, losses can be mitigated with longer
    surveys or larger apertures. We found that FoR trends did not significantly
    change with aperture size, but that a small FoR results in greater yield
    losses for shorter surveys and higher $N_\text{char}$ requirements.
    \item \textbf{Characterizations required ($N_\text{char}$):} We found that
    $N_\text{char}$ has a large impact on yield, with each additional
    characterization costing $\sim22\%$ of the mission's yield. If a higher
    $N_\text{char}$ is required, helpful trades include longer
    $T_\text{survey}$, wider FoR, or larger $D_\text{insc}$. This may also
    suggest significant benefits for designing a mission that can perform
    spectral characterizations in parallel during an observation.     
    \item \textbf{Survey duration ($T_\text{survey}$):} Extending the survey
    time always benefits yield, but it
    disproportionately benefits missions that require a higher $N_\text{char}$.
    Additional time also compensates for narrower FoR
    or smaller apertures by providing more observing windows and
    time for follow-up observations.
\end{itemize}

These trends suggest multiple viable design approaches for reaching HWO's goal
of 25 EECs. For example, a larger mirror can enable more stringent
characterization or tolerate a smaller FoR, whereas a shorter mission may
require relaxing $N_\text{char}$ or increasing FoR to preserve characterized
EEC. The appropriate mix depends on program-level constraints. The key will be
to consider the full trade space and potentially preserve at least one
engineering avenue (survey time, FoR, or aperture) to relieve the pressure
introduced if many characterizations are determined to be a necessary science
requirement.

\subsection{Limitations and Future Work}

As yield studies attempt to simulate the dynamics of a \rev{complex} future
mission with significant astrophysical and operational unknowns, there
is a nearly endless set of potential algorithmic and modeling improvements.
Important astrophysical assumptions include idealized detectors, a fixed
exozodiacal level (3 $n_\text{EZ}$), and simplified orbital-uncertainty
evolution rather than full orbit fitting. Operationally, the scheduler
design and configuration (e.g., $P_\text{det}$ threshold, 250-day look-ahead, 15-day
minimum revisit, 25-visit cap) significantly influences yield and we expect
that there are optimization opportunities to better balance discovery
and characterization. Our primary goals for future work are:
\begin{enumerate}
    \item Create a unified cost function that can balance discovery
    and characterization to eliminate the need for many heuristics used by the
    current scheduler.
    \item Implement, or better approximate, Bayesian orbit fitting to update
    $\mathcal{O}_p$ from evidence and benchmark yields against the current
    shrinkage heuristic.
    \item Expand the characterization model beyond a single-band IFS at
    910\,nm to multi-band decision trees (UV/VIS/NIR) and test how much that
    logic affects the cost of higher $N_\text{char}$.
    \item Replace the idealized, noiseless detector with realistic detector
    models including read noise, dark current, and CIC.
    \item Replace fixed exozodi ($n_\text{EZ}=3$) with randomly drawn values and explore
    adding logic to include short observations that characterize the exozodi of a system
    before searching for EECs.
    \item Perform side-by-side comparisons with AYO to cross-validate FoR,
    aperture, and observing-time trends, and reconcile dynamic-vs-probabilistic
    yields.
    \item Include observing blocks dedicated to general astrophysics to understand
    how they affect the EEC survey's scheduling efficiency.
\end{enumerate}

These additions will improve physical realism, better represent operational
constraints, and provide stronger validation of the scheduler and yield trends.

\subsection*{Disclosures}
The authors declare there are no financial interests, commercial affiliations,
or other potential conflicts of interest that have influenced the objectivity
of this research or the writing of this paper.

\subsection*{Code and Data Availability}

All software used in this work is open source, pip installable, and publicly available:
\begin{itemize}
    \item EXOSIMS (v3.4.1): \url{https://github.com/dsavransky/EXOSIMS}
    \item EXOSIMS-Plugins (v1.0.0): \url{https://github.com/CoreySpohn/exosims-plugins}
    \item orbix (v0.1.0): \url{https://github.com/CoreySpohn/orbix}
\end{itemize}
Data products and analysis scripts that support the findings of this study are
available from the corresponding author upon reasonable request. At the time of
writing, the EXOSIMS version necessary to run the analysis is hosted on the
author's fork and is in the process of being merged into the main EXOSIMS
repository. When all necessary changes are merged and included in an EXOSIMS
release the EXOSIMS-Plugins library will be updated to require that release.

\subsection* {Acknowledgments}
CS's and NL’s research was supported by an appointment to the NASA Postdoctoral
Program at the NASA Goddard Space Flight Center, administered by Oak Ridge
Associated Universities under contract with NASA. A portion of this work was
supported by NASA Goddard’s Sellers Exoplanet Environments Collaboration (SEEC).
This project was partially supported by the NASA HQ-directed ExoSpec work
package under the Internal Scientist Funding Model (ISFM). DS acknowledges
support by the National Aeronautics and Space Administration under Grant No.
80NSSC24K0139 issued through the ROSES 2022: Astrophysics Decadal Survey
Precursor Science program. Any opinions, findings, conclusions or
recommendations expressed in this material are those of the authors and do not
necessarily reflect the views of NASA.

\appendix
\crefalias{section}{appendix}

\section{\rev{Example Simulation Outputs}}
\phantomsection\label{sec:example_schedule}

\begin{figure}[t]
    \centering
    \includegraphics[width=1.0\linewidth]{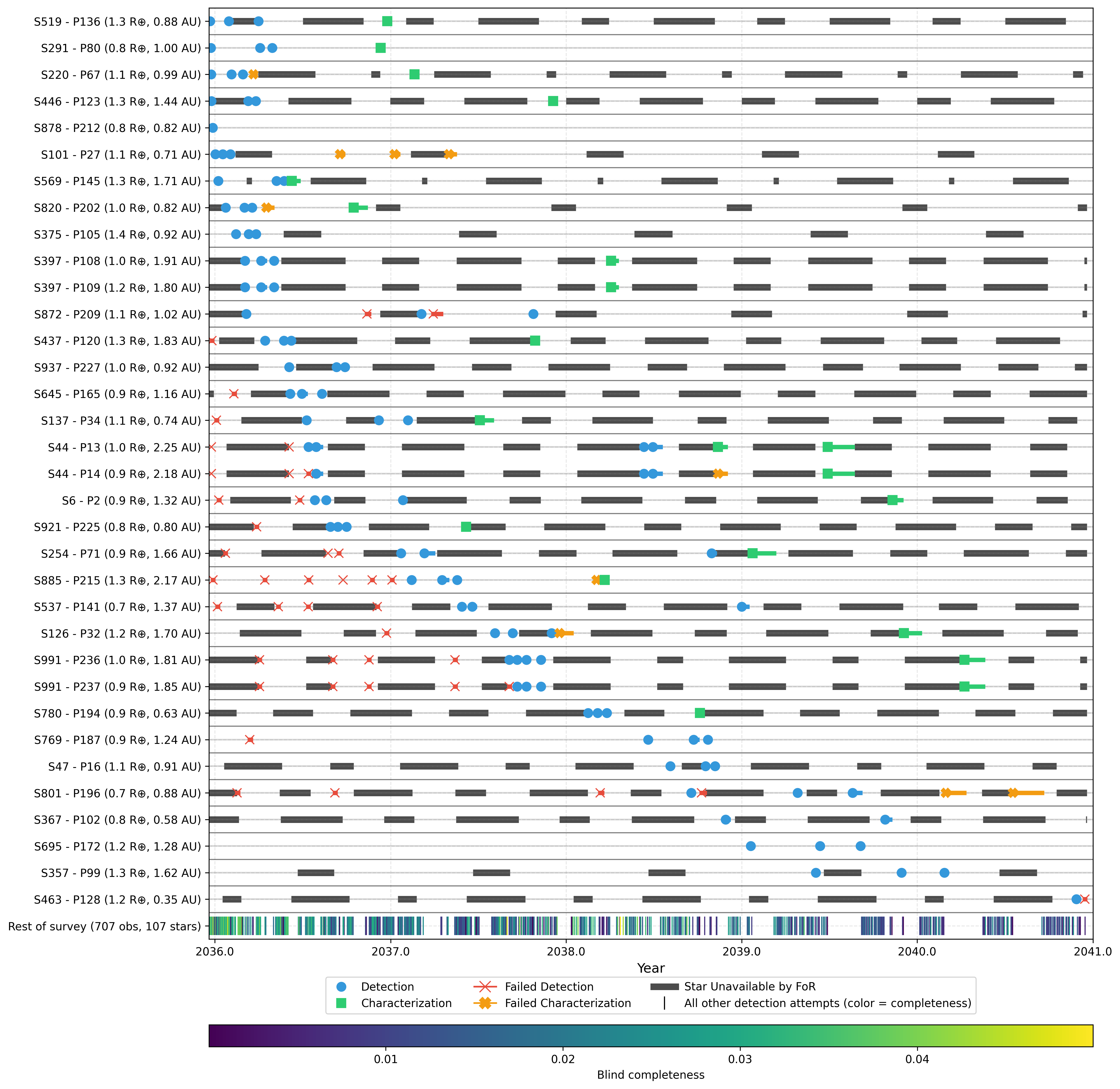}
    \caption{Example mission schedule for a 5 year mission, a 6.5\,m inscribed
    aperture, one required characterization, and FoR of $75\degree$. The
    mission detected 34 EECs and characterized 20 of them. The bottom row of the
    figure shows all blind observations of stars without a detected EEC, with
    each observation's opacity representing the completeness of the
    observation. All other rows show observations of an EEC that was detected
    during the course of the mission. The black bars indicate that the star was
    blocked by the FoR constraint. Each detection and characterization
    observation includes a line showing the duration of the observation.
    }
    \label{fig:mission_schedule}
\end{figure}

\Cref{fig:mission_schedule} presents an example mission schedule for a 5-year
mission with $D_\text{insc}=6.5$\,m, $N_\text{char}=1$, and FoR of $75\degree$.
In this realization the survey detects 34 EECs and completes 20 spectroscopic
characterizations. All rows except the bottom one correspond to an individual
EEC candidate, showing all attempted detection and characterization observations. The bottom
row represents all other observations, which are all blind search observations
made of stars that never resulted in an EEC detection. The thick black bars denote intervals
when the host star is outside the instantaneous FoR and therefore unobservable.

\begin{figure}[t]
    \centering
    \includegraphics[width=0.75\linewidth]{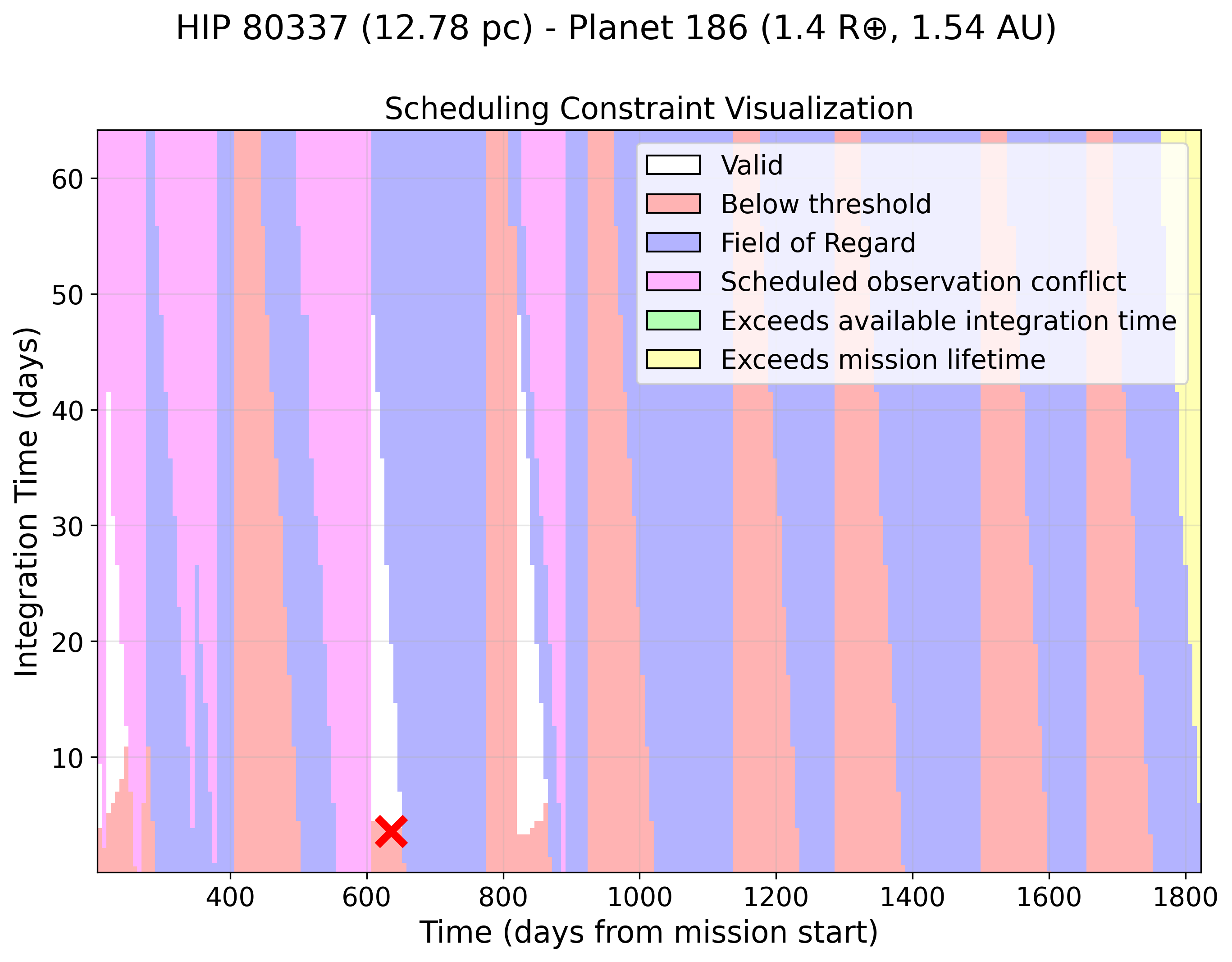}
    \caption{\rev{Scheduling constraint map for a single planet's observation. Each point on the grid represents a
    candidate observation (epoch and integration time) and shows the constraint that would prevent
    scheduling at that time: field of regard exclusion (blue), probability of
    detection below threshold (red), conflict with an already-scheduled
    observation (pink), integration time exceeding the available time budget (green),
    or exceeding the mission lifetime (yellow). White regions are valid
    observations and the red marker indicates the epoch at which this
    planet was scheduled to be observed.}}
    \label{fig:constraint_map}
\end{figure}

\rev{\Cref{fig:constraint_map} visualizes the scheduling constraints for a
single planet observation scheduling attempt. The figure makes explicit where
each constraint eliminates candidate observation windows. Notably, the FoR exclusion
in blue and the $P_\text{det}$ threshold in red combine to significantly
restrict the available observations.}

\section{High-Performance Orbital Mechanics with \texttt{orbix}}
\phantomsection\label{sec:orbix}

The dynamic, probability-driven scheduling strategy employed in this study
requires intensive computational resources. Optimizing observations in
the scheduling loop demands the ability to rapidly quantify orbital uncertainty and
forecast detection probabilities. This involves the propagation and evaluation
of millions of potential orbits throughout the simulation. Previous approaches
to calculate completeness and probability of detection were unoptimized and
posed significant computational bottlenecks.

To address this challenge, we developed \texttt{orbix}, a new open-source
Python library specifically designed for high-performance astrodynamics and
probabilistic calculations tailored for exoplanet mission simulation.
\texttt{orbix} is built using the JAX framework\cite{jax2018github} and the
Equinox library\cite{kidger_equinox_2021},
enabling Just-In-Time (JIT) compilation and
automatic vectorization (e.g., \texttt{vmap}). This architecture allows the
entire computational pipeline, from solving Kepler's equation to evaluating
instrument sensitivity, to be compiled into optimized kernels that execute
efficiently on GPUs/TPUs or in parallel on CPUs.

In this work, \texttt{orbix} provides the computational backbone for the new
EXOSIMS-plugins modules \texttt{OrbixScheduler}, \texttt{OrbixCompleteness}, and
\texttt{OrbixUniverse}. Note that \texttt{orbix} is currently in an early stage
of development. While its propagation and probabilistic capabilities are robust
and utilized here, development is ongoing to incorporate orbit fitting into
EXOSIMS simulations. A dedicated publication detailing the \texttt{orbix}
architecture and performance benchmarks is planned following the completion of
these features and more extensive benchmarking.

\subsection{Vectorized Orbit Propagation}

The primary computational bottleneck for orbit propagation is solving Kepler's
equation ($M = E - e \sin(E)$) to determine the eccentric anomaly ($E$) from
the mean anomaly ($M$) and eccentricity ($e$). \texttt{orbix} implements two
distinct JAX-based strategies for this:

\begin{enumerate}
    \item \textbf{Direct Solution:} For scenarios requiring high precision,
          \texttt{orbix} utilizes a JIT-compatible, vectorized adaptation of
            the \texttt{orvara} method~\cite{brandt_orvara_2021}.
    \item \textbf{Grid-Based Interpolation:} For massive ensemble propagation
            where precision is less important than speed (e.g., $N>10^5$
            orbits), \texttt{orbix} employs a grid-based approximation method.
            A 2D grid of $E$, $\sin(E)$, and $\cos(E)$ values is pre-computed
            across the $(M, e)$ space. JIT-compiled bilinear interpolation
            functions, leveraging efficient memory access patterns, are then
            used to rapidly estimate $E$ and the necessary trigonometric
            functions. This approach avoids branching at runtime and is easily
            vectorized, offering significant speedups (typically 3-5x on a
            Macbook Pro with an M2 CPU) over iterative solvers when evaluating many
            orbits at thousands of epochs simultaneously. In many scenarios only
            $\sin(E)$ and $\cos(E)$ are required, further reducing the
            computational overhead.
\end{enumerate}

\subsection{Probabilistic Calculations and Instrument Interface}

\texttt{orbix} facilitates the rapid interface between propagated orbital
ensembles and the pre-computed instrument sensitivity grids
($\Delta\text{mag}_0$). The library defines JIT-compatible structures (using
the Equinox library\cite{kidger_equinox_2021}) to manage these multi-dimensional sensitivity grids.
\texttt{orbix} performs the following key calculations in a vectorized manner:

\begin{enumerate}
    \item \textbf{Geometric Calculations:} For an ensemble of orbits at a given
        epoch, \texttt{orbix} calculates the apparent angular separation and
        the planet-star magnitude difference ($\Delta\text{mag}$).
    \item \textbf{Sensitivity Interpolation:} It pre-computes and performs rapid, vectorized
        multi-dimensional interpolation over the $\Delta\text{mag}_0$($t_\text{int}$, $\alpha$, $Z$, $\kappa_{\text{EZ}}$) grid
        to determine the instrument sensitivity for the conditions of each orbit.
    \item \textbf{Completeness:} \texttt{orbix} calculates and updates
        the completeness by applying the detection mask to the
        ensemble, which is used to optimize target selection during blind
        searches.
    \item \textbf{Probability of Detection ($P_\text{det}$):} By comparing the
        calculated $\Delta\text{mag}$ with the interpolated
        $\Delta\text{mag}_0$ for every orbit, a detection mask is generated.
        The $P_\text{det}$ is calculated as the mean of this mask, representing the fraction of orbits
        that are geometrically accessible (within IWA/OWA) and sufficiently
        bright to be detected.
\end{enumerate}

This vectorized approach to calculating probabilistic metrics makes the
``in-the-loop'' optimization used by the scheduler computationally feasible.
For all compiled JAX calculations involving integration times, we restrict the
allowed values to a fixed grid of 50 logarithmically spaced times between 1
hour and 60 days. Fixing the grid in this fashion keeps the array shapes
constant, allowing JAX to compile a single kernel that can be reused for the
majority of the calculations, reducing the computation and memory overhead. We
found that the log spaced grid of 50 integration times resolves the large
dynamic range of integration times without materially impacting scheduling
decisions.


\bibliography{zotero.bib}   
\bibliographystyle{spiejour}   


\vspace{2ex}\noindent\textbf{Corey Spohn} is a NASA Postdoctoral Program Fellow
at the NASA Goddard Space Flight Center. His research focuses on optimizing the
design and operations of future direct imaging missions, specifically the
Habitable Worlds Observatory (HWO). He specializes in exoplanet yield modeling
and the development of high-performance software for astrodynamics and mission
simulation. He received his Ph.D. in Aerospace Engineering from Cornell
University and bachelor's degrees in Physics and Engineering from Virginia Tech.

\vspace{1ex}
\noindent Biographies and photographs of the other authors are not available.

\listoffigures
\listoftables

\end{spacing}
\end{document}